\documentclass{ifacconf}

\usepackage{graphicx}      
\usepackage{natbib}        

\usepackage{color}
\usepackage{float}
\usepackage{mathrsfs}
\usepackage{amsmath}
\usepackage{amssymb}
\usepackage{graphicx}

\usepackage{stmaryrd}

\begin{document}

\begin{frontmatter}

\title{Identification of complex network topologies through delayed mutual information} 

\author[First,Second]{Pierre-Alain Toupance} 
\author[First]{Bastien Chopard} 
\author[Second]{Laurent Lefevre}

\address[First]{University of Geneva,  Geneva, Switzerland \\ bastien.chopard@unige.ch}
\address[Second]{Univ. Grenoble Alpes, Grenoble INP, LCIS, Valence, France  \\ \{pierre-alain.toupance,laurent.lefevre\}@lcis.grenoble-inp.fr}

\begin{abstract}                
The definitions of delayed mutual information and multi-information
are recalled.  It is shown how the delayed mutual information may be
used to reconstruct the interaction topology resulting from some
unknown scale-free graph with its associated local dynamics. Delayed
mutual information is also used  to solve the community detection
problem. A probabilistic voter model defined on a scale-free graph is
used throughout the paper as an illustrative example.
\end{abstract}


\end{frontmatter}

\section{Introduction}

In \cite{conant1976}, the author already pointed out the interest of
information theoretic approaches for the analysis of complex dynamical
systems. He specifically highlighted their additive complexity which
make them attractive for large scale systems. These approaches are
based on the computation and analysis of information flows (and
losses) between subsystems and components of the complex system.
Therefore it seems quite logical to use them for the analysis of
control systems topologies, with applications in mind to coarse
graining or partitioning (decomposition) problems, for
instance. Delayed mutual information was introduced in
\cite{schreiber2000} - with appropriate conditioning of transition
probabilities - to distinguish driving and responding elements through
the analysis of the correlation between two stochastic
signals. Introducing the time delay and detecting asymmetry in the
interaction of subsystems allows one to distinguish information that
is actually exchanged between two subsystems from shared information
due to common history and input signals.

The structural analysis of complex systems dynamics and input-output
properties has a long history. Approaches have been developed which
make use of interconnection graph or "inference diagram", describing
the existing (analytical) relations between a priori given input,
state and output variables (\cite{Lin74}; \cite{Siljak11}. Such
approaches give structural results on controllability and
observability, together with efficient graph algorithms. Some recent
results on structural controllability and observability using this
approach are presented in \cite{Liu11,Liu13}. They require a priori
knowledge of the system interconnection topology. When trying to
isolate the most influential or measurable nodes in some complex
system, this information is often missing or incomplete. Besides,
these results on controllability and observability only conclude to
some existing causal relation between the considered sets of
variables, but not how much the dynamics of a node may be measured or
controlled from another node.

Therefore we proposed in \cite{Toupance19} the use of delayed mutual
and multi-informations to analyze the most influential components in a
complex system with no a priori knowledge on the interconnection
topology. This approach is non-intrusive in the sense that it may be
performed by simply sampling the  state dynamics, even if the
underlying dynamics is unknown. We proved - on the example of the
so-called voter model - that the nodes may be ranked according to
their influence (the impact on the average opinion of the entire
group) by monitoring the time-delayed multi-information and that this
ranking closely relates to controllability/observability grammians
singular values.

In this paper we investigate how this delayed mutual information
approach may be used to reconstruct the interconnection topology (for
instance the incidence matrix when the dynamics is defined on a
graph). Our goal is to develop an approach which is only based on the
sampling of the state dynamics and could be further applied online,
for instance with a moving time frame for the computation of the
mutual information. Ideally the approach should be effective when
changes occur in the topology, making us able to detect theses changes
and reconstruct quickly the new system topology. We will consider
again - as an illustration example - a probabilistic voter model where
the vote dynamics is defined on a scale-free graph representing
somehow the influence between agents in a social network. The
quantitative nature of the mutual information and multi-information
suggests a new way to detect groups or communities in the complex
system, based on the information flux between these communities. We
will show how this partitioning algorithm works on the example of the
voter model and will suggest that it could be an alternative approach
either for coarse graining or partitioning, or for estimation, control
or diagnosis purpose (see for instance \cite{ocampo2011}).

The paper is organized as follows: section~\ref{sect:voter} introduces
the voter model which will be used throughout the paper (subsection
~\ref{subsect:model}) and the metrics from the theory of information
that we will use (subsection \ref{subsect:mutualinfo}), together with
a summary of results previously obtained when using these metrics to
measure the relative influence of the agents in the voter model. Our
main contributions are presented in section~\ref{sect:topology} where
it is shown that time-delayed multi-information and mutual information
allow us to reconstruct the interconnection topology of a complex system
(subsection ~\ref{subsect:adjancymatrix}) and partition the graph into
communities which are defined from the information flows (subsection
~\ref{subsect:community}).

\section{The voter model}\label{sect:voter}

\subsection{Description of the model}\label{subsect:model}
Simple models that abstracts the process of opinion formation have
been proposed by many
researchers~\cite{RevModPhys.81.591,BC-galam:98}.  The version we
consider here is an agent-based model defined on a graph of arbitrary
topology, whether directed or not.

A binary agent occupies each node of the network. The dynamics is
specified by assuming that each agent $i$ looks at every other agent
in its neighborhood, and counts the percentage $\rho_i$ of those which
are in the state $+1$ (in case an agent is linked to itself, it
obviously belongs to its own neighborhood). A function $f$ is
specified such that $0\le f(\rho_i)\le 1$ gives the probability for
agent $i$ to be in state $+1$ at the next iteration. For instance, if
$f$ would be chosen as $f(\rho) = \rho$, an agent for which all neighbors
are in state $+1$ will turn into state $+1$ with certainty.  The
update is performed synchronously over all $n$ agents.

Formally, the dynamics of the voter model can be express as
\begin{equation}
s_i(t+1)=\left\{ \begin{array}{cc}
                  1 & \mbox{with probability $f(\rho_i(t))$} \\
                  0 & \mbox{with probability $1-f(\rho_i(t))$}\\ 
                 \end{array}
\right.
\label{eq:votermodel}
\end{equation}
where $s_i(t)\in\{0,1\}$ is the state of agent $i$ at iteration $t$,
and 
\begin{equation}
\rho_i(t)={1\over |N_i|}\sum_{j\in N_i} s_j(t).
\end{equation}
The set $N_i$ is the set of agents $j$ that are neighbors of agent $i$,
as specified by the network topology.

The global density of all $n$ agents with opinion $1$ is obviously
obtained as
\begin{equation}
  \rho(t)={1\over n}\sum_{i=1}^n s_i(t)
  \label{eq:rho}
\end{equation}

In what follows, we will use a particular function $f$, 
\begin{equation}
 f(\rho)= (1- \epsilon) \rho + \epsilon (1-\rho)=(1-2\epsilon)\rho + \epsilon 
\end{equation}
The quantity $0\le\epsilon\le1/2$ is called the noise. It reflects the
probability to take a decision different from that of the
neighborhood.


To illustrate the behavior of this model, we consider a random
scale-free graph $G$, as simple instance of a social
network~(\cite{Barabasi2000}). We use the algorithm  of B{\'e}la Bollob{\'a}s (\cite{Bollobas2003}) to generate this graph.

\begin{figure}
\begin{center}
\includegraphics[width=.5\textwidth]{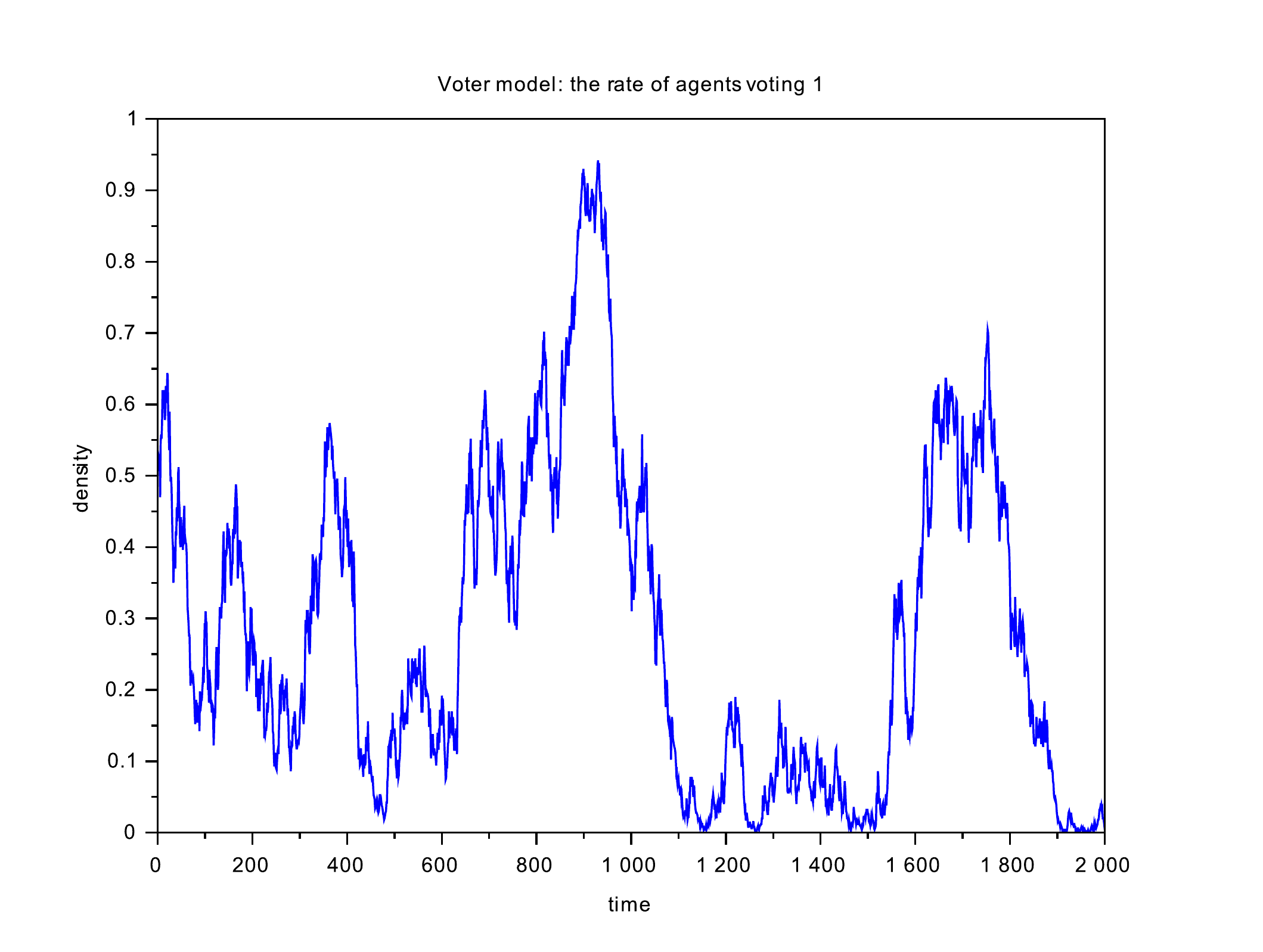}
\end{center}
\caption{Time evolution of the density of opinion 1 with
  noise $\epsilon=0,001$ and $n=200$ agents connected through a
  scale-free network.}
\label{fig:simu1}
\end{figure}

Figure~\ref{fig:simu1} shows the corresponding density of agents with
opinion 1, as a function of time. We can see that there is a lot of
fluctuations due to the fact that states ``all 0's'' or ``all 1's'' are no
longer absorbing states when $\epsilon\ne0$.

\subsection{Delay Mutual and multi-information}\label{subsect:mutualinfo}
Let us consider a set of random variables $X_i(t)$ associated with
each agent $i$, taking their values in a set $A$. For instance,
$X_i(t)=s_i(t)$ would be the opinion of agent $i$ at iteration $t$.

Since we want to assess the temporal causality, we measure the
influence of the vote of agent $i$ at time $t$ on the vote of agent
$j$ at time $t+\tau$, we define the $\tau$-delayed mutual information
$w_{i,j}$ as
\begin{eqnarray}
\omega_{i,j}(t,\tau)&=&I(X_i(t),X_j(t+\tau))  \\
&=&
\sum_{(x,y) \in A^2} p_{xy} \log \left( {p_{xy} \over p_x p_y} \right)
\label{info_mutual}
\end{eqnarray}
with 
$ p_{xy}= \mathbb{P} (X_i(t) =x,X_j(t+\tau)=y)$ and \\
  $p_x = \mathbb{P} (X_i(t) =x) \text{ and } p_y= \mathbb{P} (X_j(t+\tau)=y)$\\

We also define the $\tau$-delayed multi-information $w_{i}$ as a
measure of the influence of agent $i$ on all the others
\begin{eqnarray}
\omega_{i}(t,\tau)&=&I(X_i(t),Y_i(t+\tau))
\label{info_multi}
\end{eqnarray}
\begin{equation}
Y_i(t+\tau) = \sum_{ k \neq i} X_k(t+\tau)
\end{equation}

These information metrics can be computed by sampling. In the sequel
we will consider $N=10^5$ instances of the system in order to perform
an ensemble average.  According to the central limit theorem, we know
that, with this number of instances, we obtain a precision of $3
\times 10^{-2}$ with a risk of $5 \%$ for the approximate values of
the probabilities that we compute.
 
\begin{figure}
\begin{center}
 \includegraphics[width=.5\textwidth]{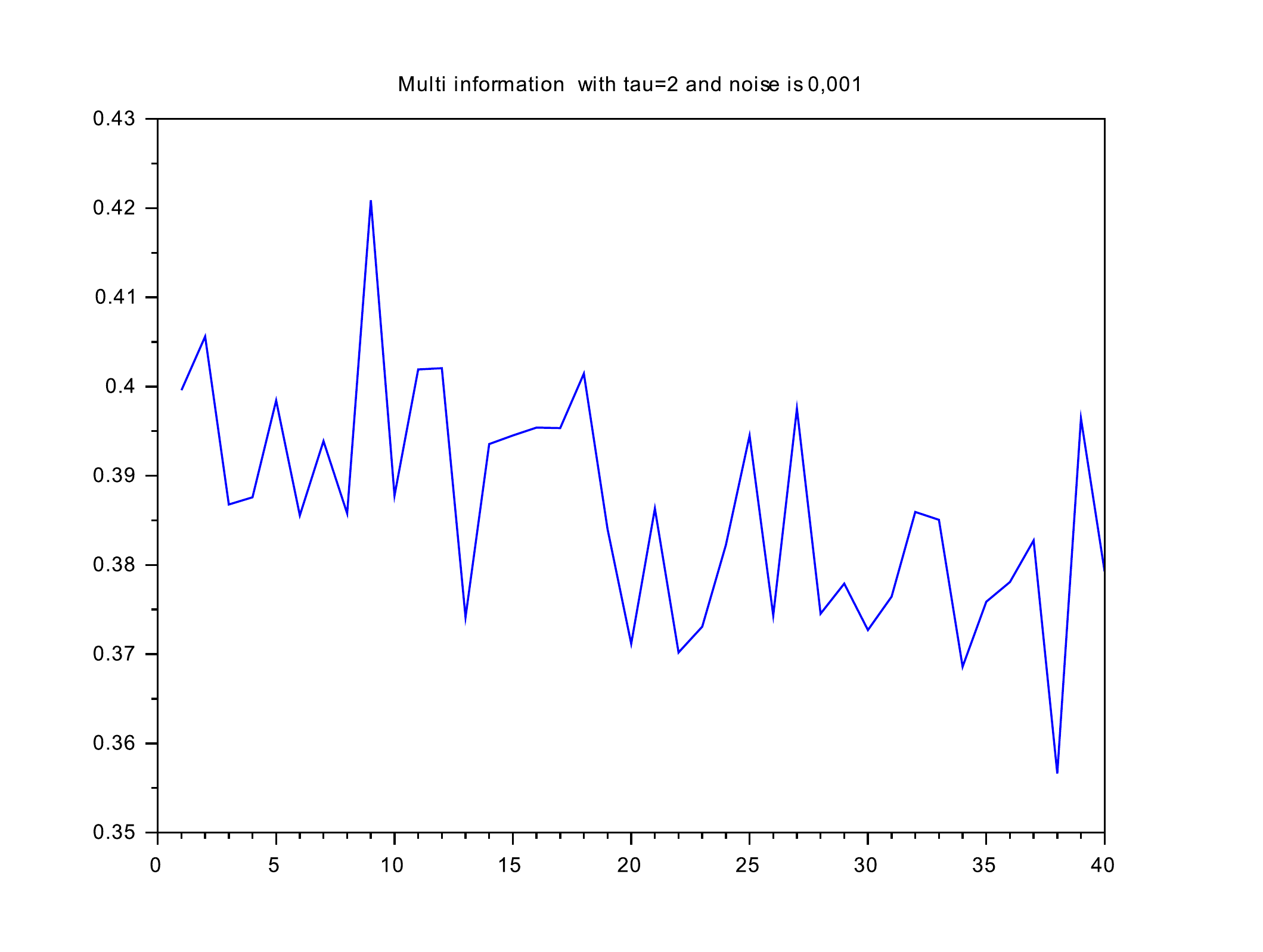}
 \end{center}
\caption{$\tau$-delayed multi-information $w_i(\tau)$ as a function of
  $i$, for a test graph $G$, similar but not identical to the graph
  shown in Fig.~\ref{fig:graph_colored}. The case show here has $n=50$
  agents and a noise level $\epsilon=0.001$.}
 \label{fig:Multi_Info}
 \end{figure}

The $\tau$-delayed multi-information can be used as a measure of the
influence of opinion of each node $i$ on the vote of the other agents.
For instance, Fig.~\ref{fig:Multi_Info} shows $\omega_i(\tau=2)$ in a
steady state, where the origin of time is arbitrary. We observe that
some agents $i$ exhibit a more pronounced peak of multi-information
towards the rest of the system, suggesting that the opinion of these
agents may affect the global opinion of all agents. Note that this
results is obtained only by probing the systems, without modifying any
of its components. For this reason, we describe this approach as
``non-intrusive''.

\subsection{Controllability and information theoretical}\label{subsect:controllability}
In the paper ``Controllability of the Voter Model : an information
theoretic approach'' (\cite{Toupance19}), we showed that the delayed
multi-information is indeed a good metrics to identify the influential
agents of the system.  To illustrate this result we compared
$w_i(\tau)$ with the impact of forcing the vote of agent $i$ to 1 at
all time. As a result of this forcing, the density of vote
\begin{equation}
  \rho(t)={1\over n}\sum_{j=1}^n s_j(t)
\end{equation}
oscillates around a value different from that observed when agent $i$
is free.  This variation of the average value of $\rho$ is defined as
the {\it influence} of agent $i$ on the whole system. This measure of
influence is ``intrusive'' as it is the result of an action on the system.

\begin{figure*}
\begin{center}
\includegraphics[width=1\textwidth]{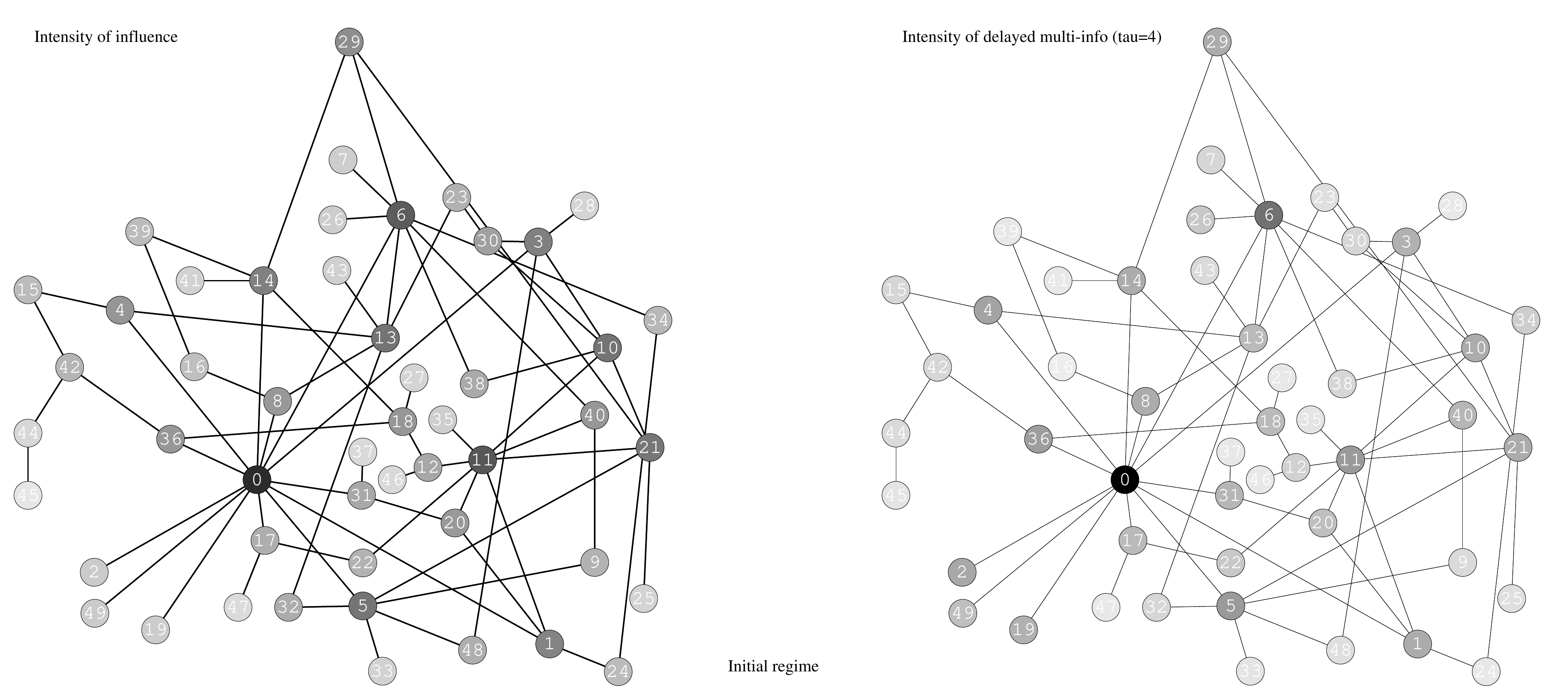}
\end{center}
\caption{Scale free graph colored as a function of the values of the
  influence (left) and the $\tau$-delayed multi-information (right),
  for $\tau=4$. The value of $\tau$ is chosen so as to match the diameter
  of the graph. In this case, the multi-information is computed in the
  transient regime that follows the initial state.}
\label{fig:graph_colored}
\end{figure*}

The two metrics (intrusive and non-intrusive) are shown in
Fig.~\ref{fig:graph_colored}, with a color representations, for the given
scale-free graph.  It shows that the multi-information (right panel)
detects correctly the influence of the nodes (left panel) since the
variation of gray levels are similar in both cases.  In this way we
can then identify, by non-intrusive observations, which agents are
those whose control will be the most influential to the system when
their vote is forced.

The strong correlation that exists between influence and the delayed
multi-information can be proven rigorously in the case of a 1D
unidirectional voting model. In \cite{Toupance19} it is shown that the
probability $\pi_i$ that agent $i$ votes $1$ when the system is
stationary, is
\begin{equation}
\pi_i = {1\over 2} + {1\over2}\exp\left[-{i\over\ell_c}\right]
\end{equation}
where $\ell_c$ is defined as 
\begin{equation}
\ell_c={1\over\ln\left({1+2\epsilon\over 1-2\epsilon}\right)}
\label{eq:control-length}
\end{equation}

\begin{figure}
  \includegraphics[width=.4\textwidth]{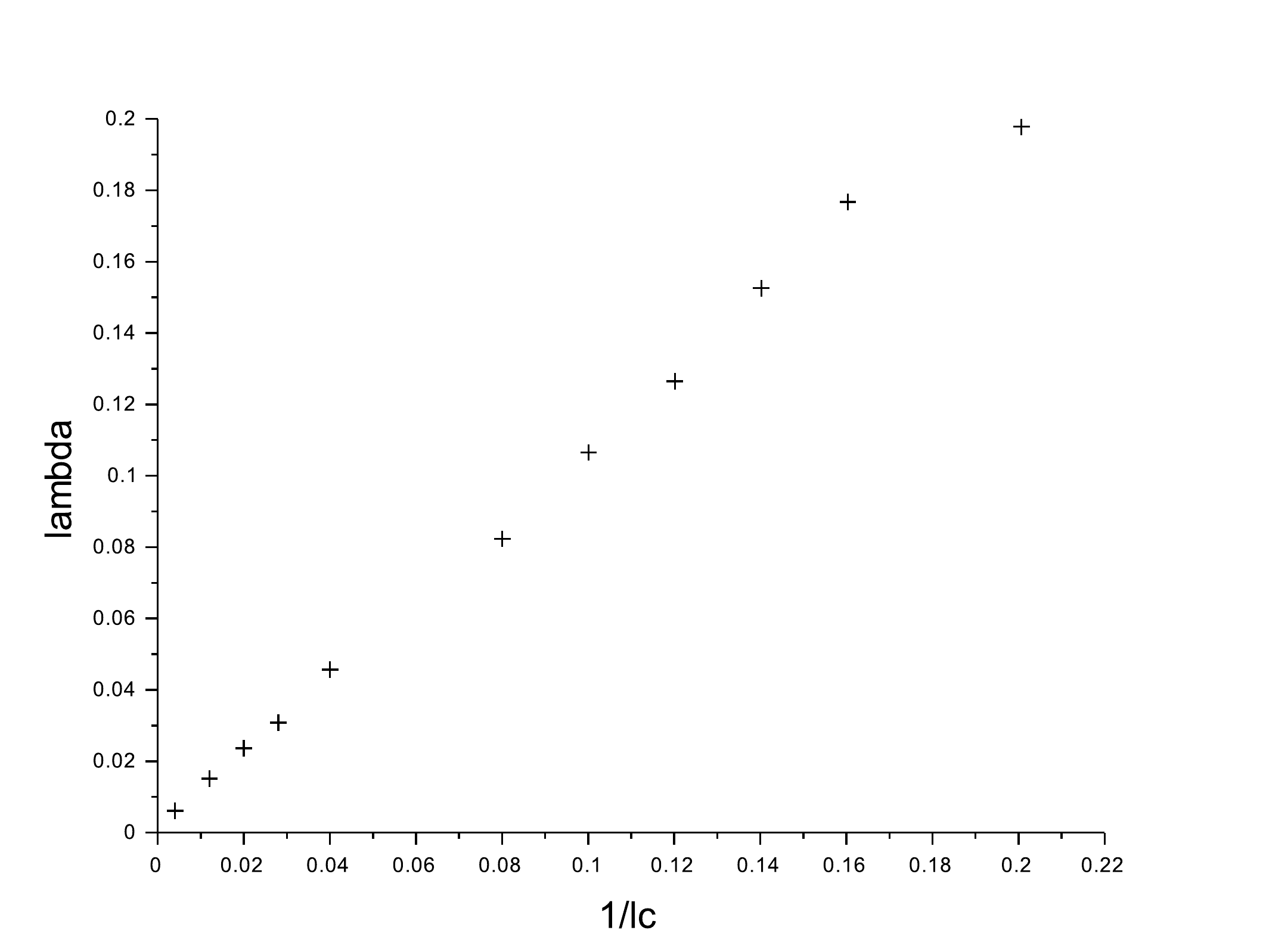}
   \label{fig:lambda-lc}
\end{figure}

We have also shown that delayed mutual information decreases
exponentially with the distance between two agents and the noise. By
simulation, we obtained that the multi-information between agents
$i$ and $j$ (where $j > i$) with a delay $\tau=j-i$ is
\[ \omega_{i,j}(j-i) =
\alpha_i \exp\left(- \lambda_i (j-i)\right)\]
where $\lambda_i$ depends on the noise level, $\epsilon$.
Figure~\ref{fig:lambda-lc} shows that $ \lambda_i $ is proportional to
$1/\ell_c$, confirming the strong link between our concept of
influence and that of time-delayed multi-information.

%

\section{Topology of the system}\label{sect:topology}
\subsection{1-delayed mutual information and adjacency matrix}\label{subsect:adjancymatrix}

After the results presented in the previous section about the link
between controllability and information theory, we are interested here
in the identification of topology of a system. The aim is to us our
information metrics to construct the interaction topology of the
unknown graph underlying the dynamics of a complex system.

\begin{figure}
\begin{center}
\includegraphics[width=8.4cm]{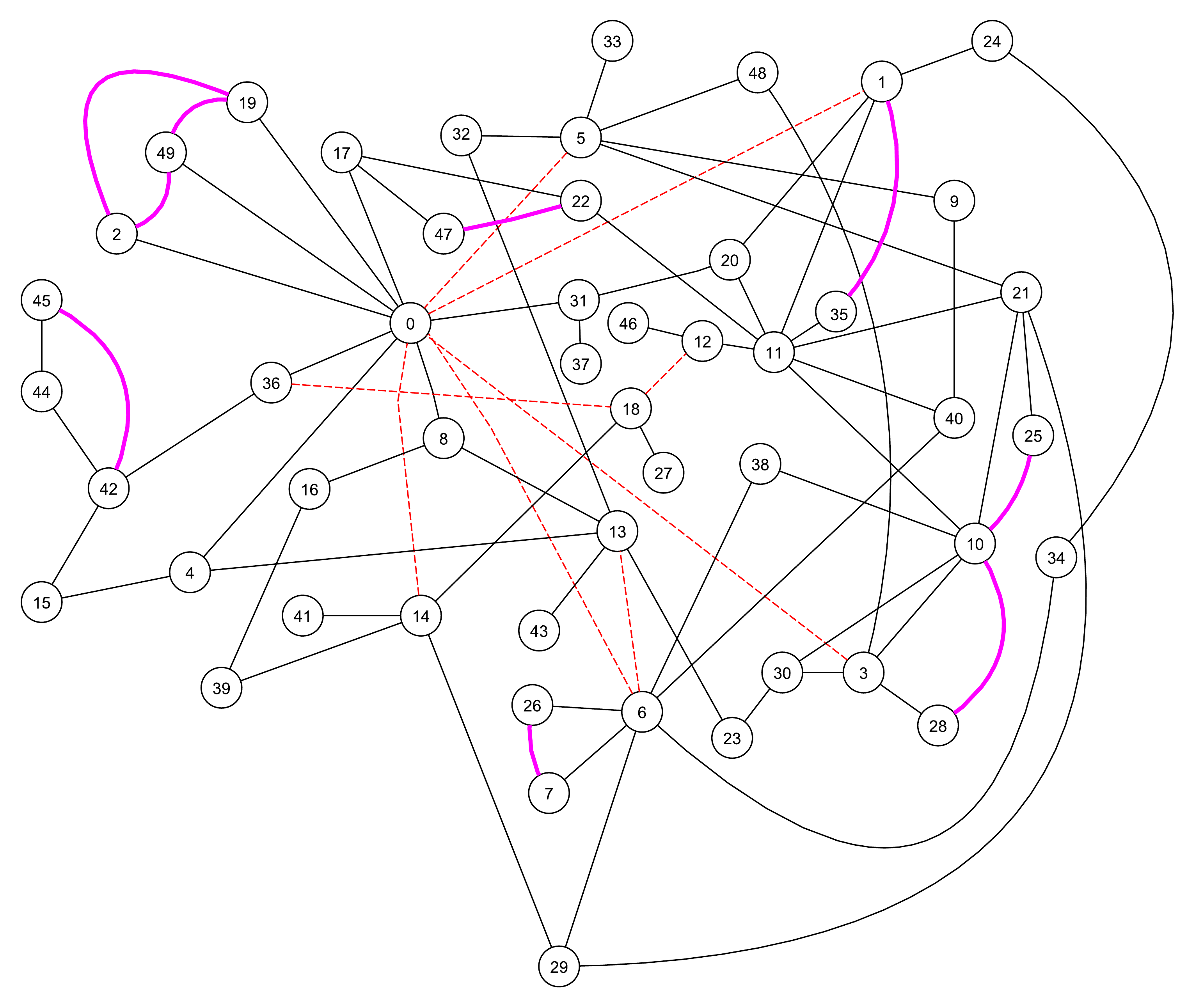}    
\caption{$1$ Delayed mutual information between agent $i=42$ and the rest of the system. Peaks are visible for the neighbors of $i$.} 
\label{fig:IM1}
\end{center}
\end{figure}

Figure~\ref{fig:IM1} shows the values of $1$-delayed mutual
information $\omega_{i,j}(1)$ between agent $i$ and all the others in the
system. These values were calculated by sampling, when the system has
reached its steady state.  In this case, the highest values of
$\omega_{i,j}(1)$ are obtained for the neighbors of agent $i$. For $i=42$
we observed a peak for agents $15$, $36$, $42$, $44$ (the agent is
neighbor of itself).

Thus, we can use the $1$-Delayed Mutual information  to get the edges of a graph. For each agent $i$, we fixed a threshold $T_i$ for $\omega_{i,j}(1)$ that indicates
that $j$ is a neighbor of $i$.  This threshold is defined as
$$T_i = \mu_i + a_i \sigma_i$$ where $ \mu_i$ is the mean of the
values of the $1$-delayed mutual information between agent $i$ and the
others agents, and $\sigma_i$ is the standard deviation of these
values.

There are two possible values for  $a_i$ according to the following criteria
\begin{itemize}
\item if agent $i$ is very influential, that is its delayed multi-information is high, then we take $a_i= 0.2$.
\item otherwise $a_i= 0.7$
\end{itemize} 
These two thresholds are motivated by the fact that influential agents usually have more neighbors than other agents .

The estimation of the adjacency matrix, denoted $M=(m_{i,j})_{1 \leqslant i,j \leqslant N}$, is defined by : 
$$m_{i,j}=\begin{cases} 1 &  \text{ if } \omega_{i,j}(1)> T_i \text{ or } \omega_{j,i}(1)> T_j  \\
                        0 & \text{ otherwise}
                        \end{cases}$$
In other words, when $\omega_{i,j}(1)> T_i$ or $\omega_{j,i}(1)> T_j$,
it is assumed that agents $i$ and $j$ are neighbors.

\def\r{r} 
The values of $a_i$ have been chosen in order to minimize the error
rate $\r$ between this matrix and the actual adjacency matrix $A$,
over several scale free graphs, by testing all possible values of $a$
from 0 to 1 with a step of $0.1$.  The error rate is defined by $\r =
{ \Delta(M,A) \over n^2}$, where $\Delta(M,A)$ is the Hamming
distance, namely the number of values that differ between $M$ and $A$.
$n$ is the number of agents in the graph.

\begin{figure}
\begin{center}
\includegraphics[width=8.4cm]{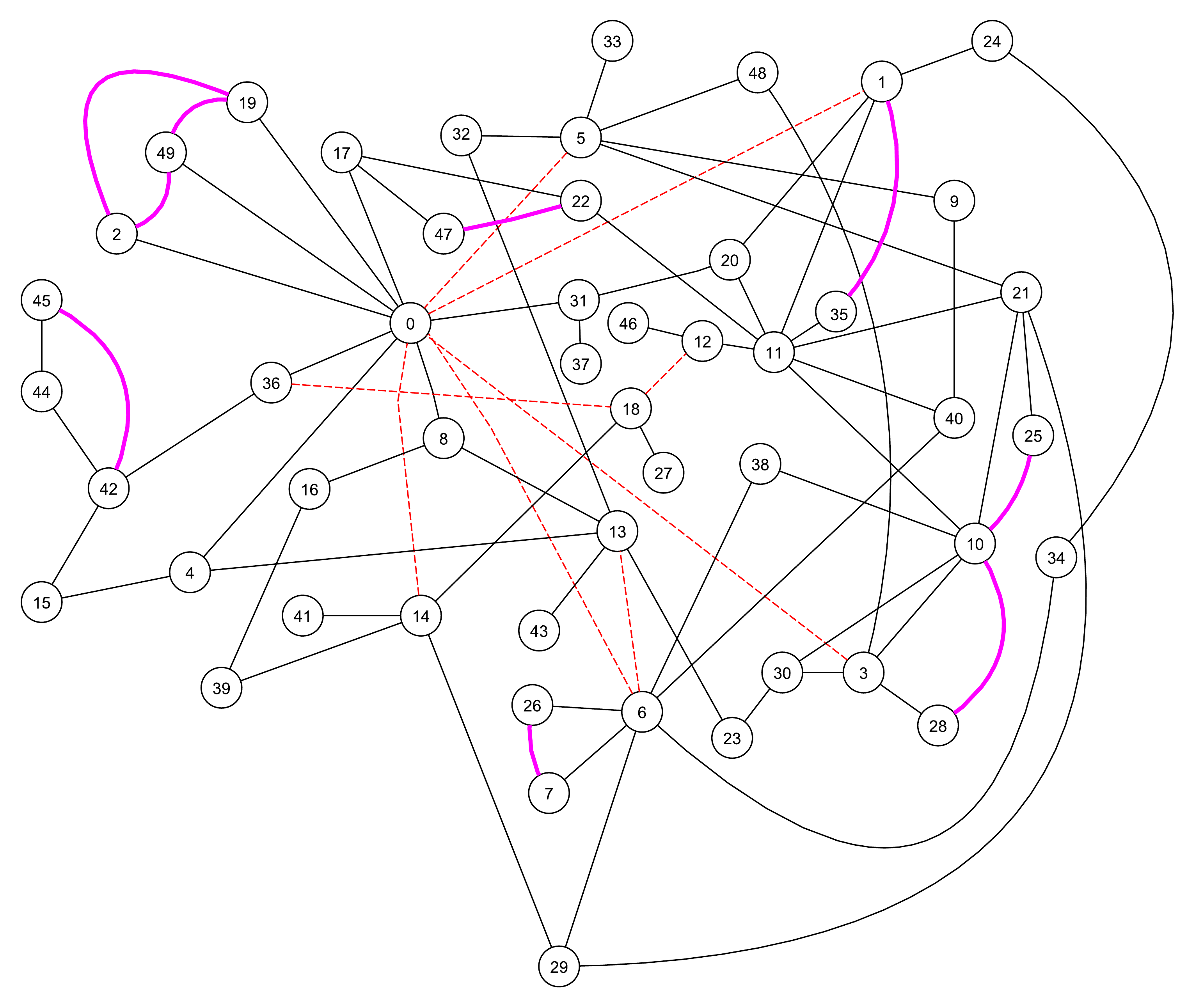}    
\caption{Graph built with $1$-delay mutual information calculated when
  the system is in a steady state. The dashed red edges are the ones
  that have not been found, and the solid pink edges are those that
  were wrongly added.}
\label{fig:top2}
\end{center}
\end{figure}

\begin{figure}
\begin{center}
\includegraphics[width=8.4cm]{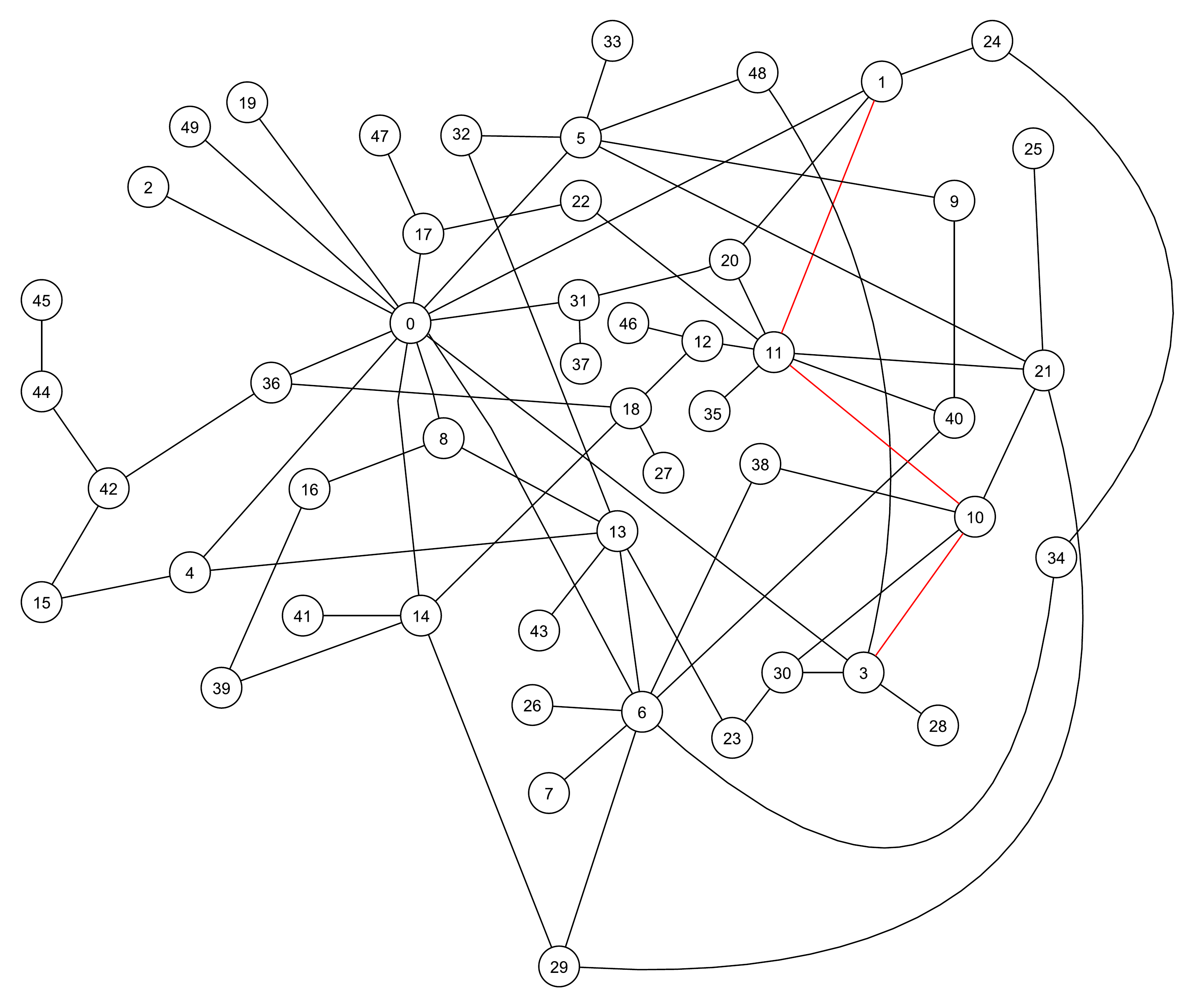}    
\caption{Graph 
built with the $1$-delayed mutual information computed when the system evolves from its initial state. The red edges are the ones that have not been found.} 
\label{fig:top1}
\end{center}
\end{figure}

For example, Fig.~\ref{fig:top2}, shows the graph that is
reconstructed by this procedure, and compares it to the original
graph. In this case, the error rate is $\r =1.3 \%$.  An even better
result is obtained when $1$-delay mutual information is computed
during the initial transient regime (see Fig.~\ref{fig:top1}). The
error rate is now $\r = 0.24 \%$. In the transient regime the results
are probably better because it really probes the direct influences.
In order to provoke such a transient regime, one may disrupt the system by temporarily increasing the
noise, while calculating the mutual information. This method has been
tested by randomly generating 20 scale free graphs. The average error
rate obtained is $0.9 \%$ and the standard deviation is $0.0026$.

\subsection{Community detection}\label{subsect:community}

The existence and structures of communities in a graph is an important
concept in the analysis of social networks.  Communities are sub-graph
with dense internal connections and sparse connections between these
sub-graphs. In our case, partitioning the agents in communities should
allow us to better control the network. In (\cite{papadopoulos2012})
the author presents different method to determine communities. In our
case, we will rather use the delayed mutual information. In a scale
free graph, we expect a seed-centric approaches for community
detection (see \cite{kanawati2014}). In the voter model these seeds
are the most influential nodes, each of them characterizing a
different community. IN the first part of this paper, we showed that
the delayed multi-information can determine these seeds. Then, to
build the corresponding community, we define the proximity $p(i,j)$
between two agents $i$ and $j$ as
\begin{equation}
p(i,j) = {1 \over r} \sum_{\tau=1}^r\omega_{i,j} (\tau)
\label{eq:proximity}
\end{equation}
where $r$ is an approximation of  radius of the graph.  \cite{bollobas2004}
proved that the diameter of a scale-free random graph is
asymptotically ${\log (n) \over \log( \log(n)) }$ where $n$ is the
size of the graph. We can choose
 $$r=\lfloor {1 \over 2} {\log (n) \over \log( \log(n))}  \rfloor$$
 this value is approximately the radius of the graph. 
 
\begin{figure}
\begin{center}
\includegraphics[width=8.4cm]{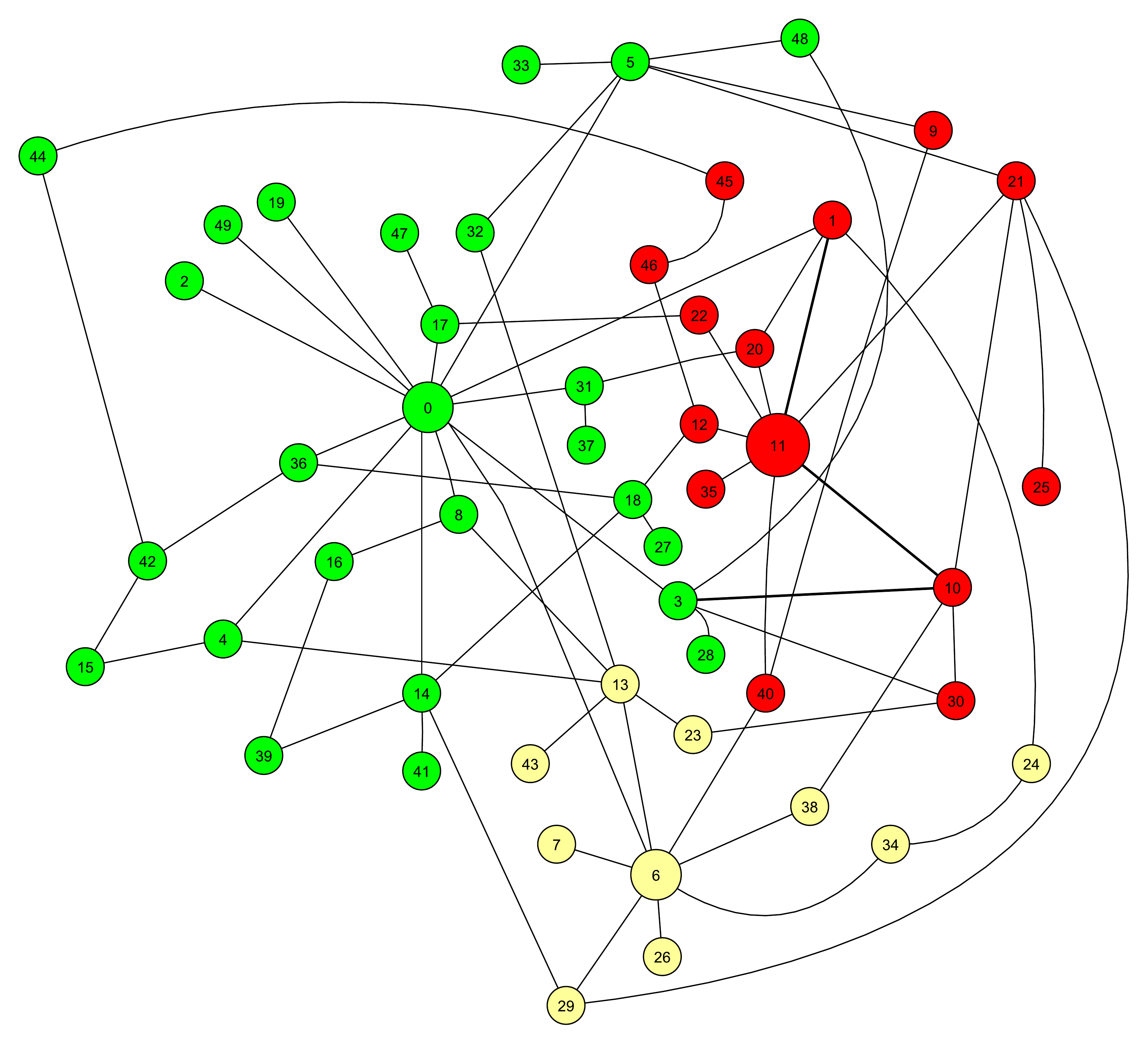}    
\caption{Partition of the graph in 3 communities.} 
\label{fig:community}
\end{center}
\end{figure}

\begin{figure}
\begin{center}
\includegraphics[width=8.4cm]{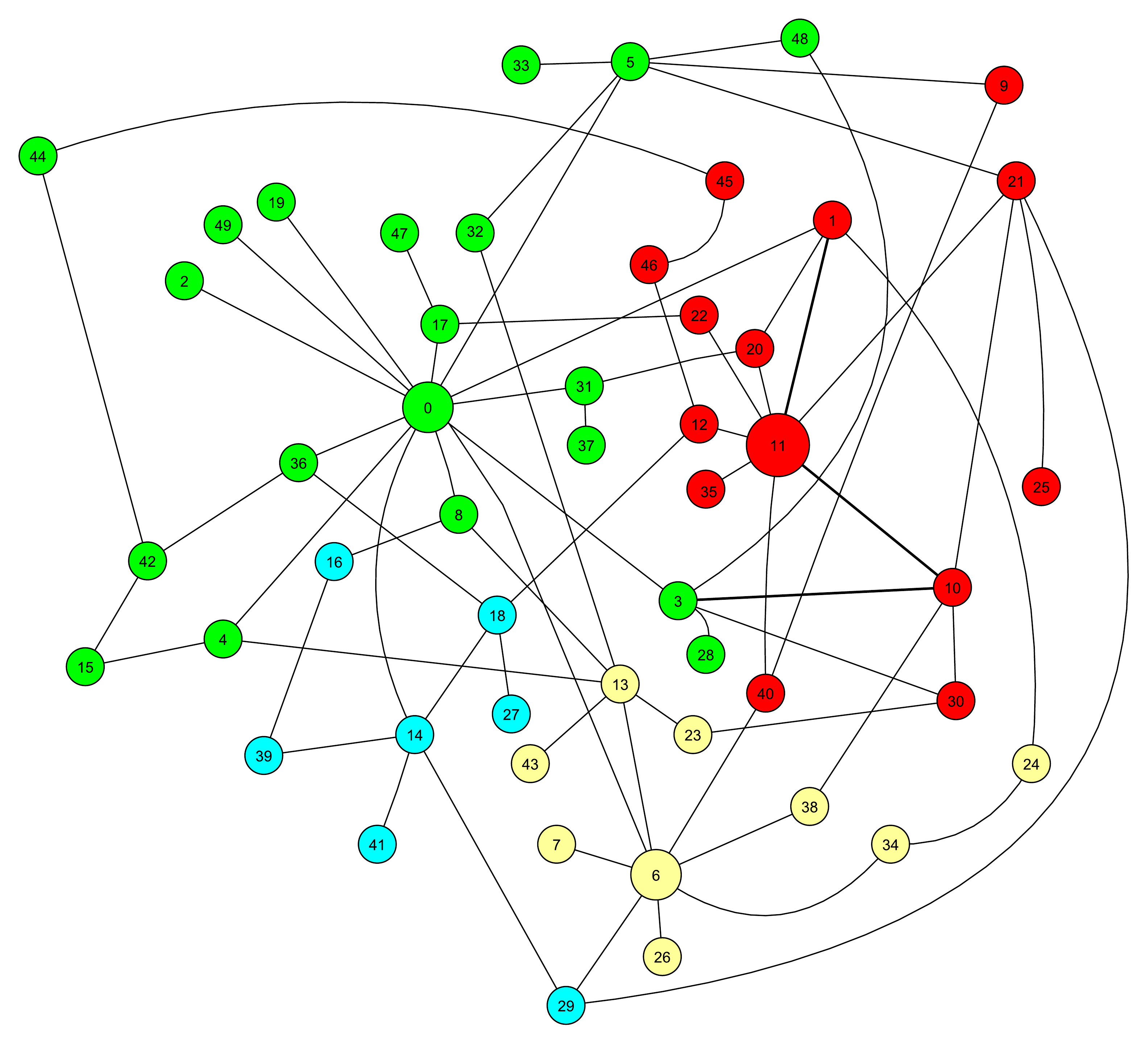}    
\caption{Partition of the graph in 4 communities.} 
\label{fig:community4}
\end{center}
\end{figure}

\begin{figure}
\begin{center}
\includegraphics[width=8.4cm]{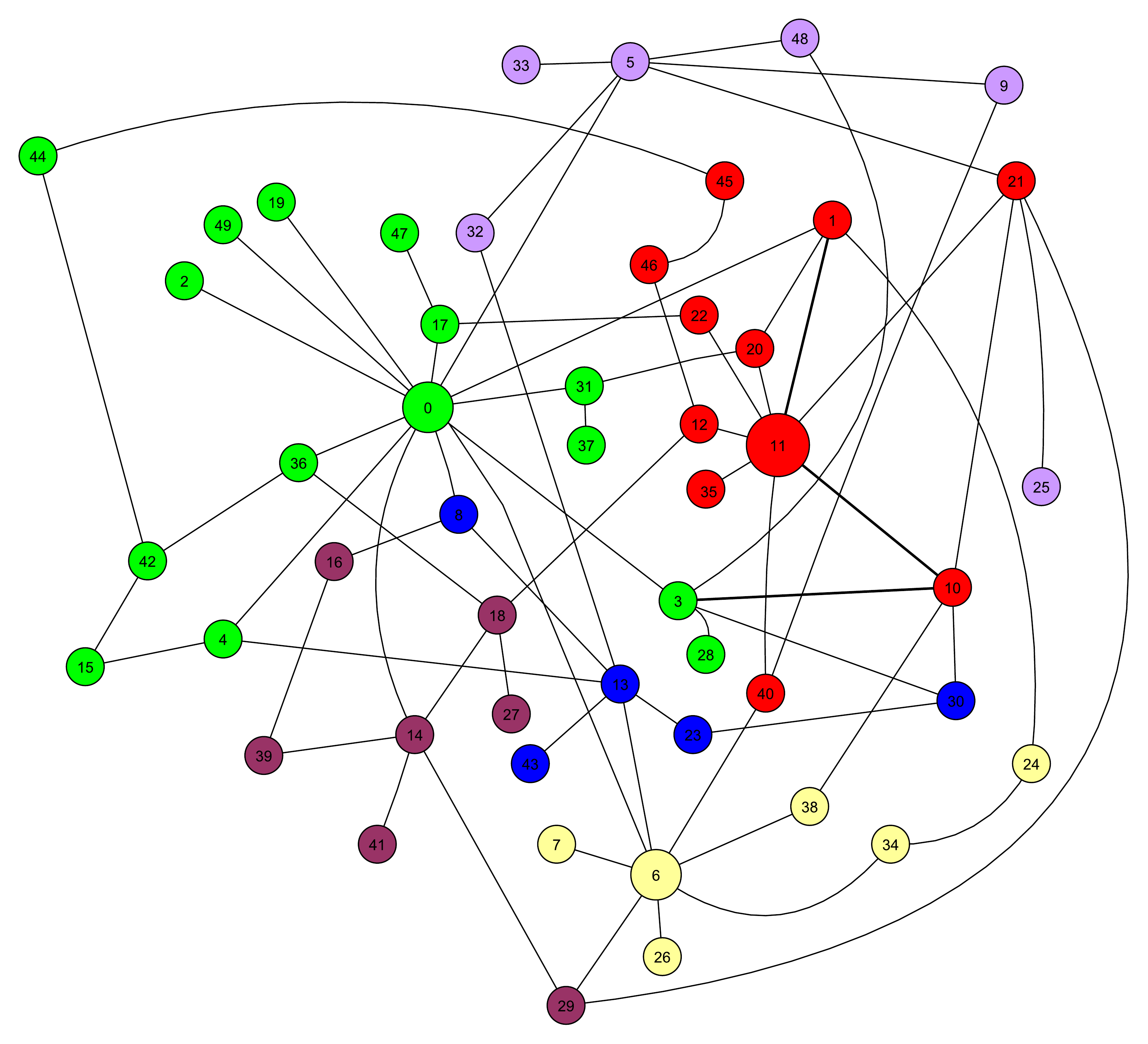}    
\caption{Partition of the graph in 6 communities.} 
\label{fig:community6}
\end{center}
\end{figure}

To determine the community of each agent, its proximity to every seed
is computed. The communities obtained with this algorithm are shown in Figs.~\ref{fig:community}, \ref{fig:community4} and
\ref{fig:community6}.

To evaluate the quality of the partitioning, we compute their
modularity $Q$ (see \cite{newman2006}) defined by :
 \begin{equation}
 Q= {1 \over 2m} \sum_{i,j} \Big( A_{i,j} - { q_i q_j \over 2m} \Big) \delta(c_i,c_j)
 \end{equation}
 where $m$ is the number of edges of the graph, $A_{i,j}$ is the
 coefficient $(i,j)$ of the adjacency matrix, $q_i$ is the degree of
 agent $i$, $c_i$ is the community of  agent $i$, and $\delta$ is
 the Kronecker symbol, we have $$\delta(c_i,c_j) = \begin{cases} 1 &
   \text{ if } i \text{ and } j \text{ are in the same community} \\ 0
   & \text{ otherwise} \end{cases}$$
 
 \begin{table}[hb]
\begin{center}
\caption{Modularity for different partitioning.}
\begin{tabular}{cccc}
Number of communities & 3 & 4 & 6 \\ \hline
$Q$ & $0.3607$ & $0.4208$ & $0.4382$ \\ \hline
\end{tabular}
\end{center}
\label{tb:modularity}
\end{table}
For our example, the coefficients of modularity are shown in
table~\ref{tb:modularity}. This indicates a good assortativity of
the proposed communities.

\section{Conclusion}
In this paper we showed that information theory is related to control
theory and further that is offers efficient tools to determine the
unknown interaction topology of a complex dynamical system. The
results are obtained using a sampling approach on a voter model
defined on a scale free graph. The key information-theoretic
quantities introduced in this work is the $\tau$-delayed mutual
information and multi-information. In addition to reconstructing the
graph topology, it can be used to determine communities that partition
the graph.

In a future work we plan to extend the present results in three
directions: (1) to detect possible changes in the graph topology over
time. This will be obtained by computing the information metrics over
sliding time windows; (2) To used the detected communities to apply
different control strategies; (3) to use community as a way to reduce
the complexity of the full system (model reduction).

\bibliography{biblio}

\begin{thebibliography}{16}
\providecommand{\natexlab}[1]{#1}
\providecommand{\url}[1]{\texttt{#1}}
\providecommand{\urlprefix}{URL }
\expandafter\ifx\csname urlstyle\endcsname\relax
  \providecommand{\doi}[1]{doi:\discretionary{}{}{}#1}\else
  \providecommand{\doi}{doi:\discretionary{}{}{}\begingroup
  \urlstyle{rm}\Url}\fi

\bibitem[{Barab{\'a}si et~al.(2000)Barab{\'a}si, Albert, and
  Jeong}]{Barabasi2000}
Barab{\'a}si, A.L., Albert, R., and Jeong, H. (2000).
\newblock Scale-free characteristics of random networks: the topology of the
  world-wide web.
\newblock \emph{Physica A: statistical mechanics and its applications},
  281(1-4), 69--77.

\bibitem[{Bollob{\'a}s and Riordan(2004)}]{bollobas2004}
Bollob{\'a}s, B. and Riordan, O. (2004).
\newblock The diameter of a scale-free random graph.
\newblock \emph{Combinatorica}, 24(1), 5--34.

\bibitem[{Bollob{\'a}s and Riordan(2003)}]{Bollobas2003}
Bollob{\'a}s, B. and Riordan, O.M. (2003).
\newblock Mathematical results on scale-free random graphs.
\newblock \emph{Handbook of graphs and networks: from the genome to the
  internet}, 1--34.

\bibitem[{Castellano et~al.(2009)Castellano, Fortunato, and
  Loreto}]{RevModPhys.81.591}
Castellano, C., Fortunato, S., and Loreto, V. (2009).
\newblock Statistical physics of social dynamics.
\newblock \emph{Rev. Mod. Phys.}, 81, 591--646.
\newblock \doi{10.1103/RevModPhys.81.591}.
\newblock \urlprefix\url{https://link.aps.org/doi/10.1103/RevModPhys.81.591}.

\bibitem[{Conant(1976)}]{conant1976}
Conant, R.C. (1976).
\newblock Laws of information which govern systems.
\newblock \emph{IEEE transactions on systems, man, and cybernetics}, (4),
  240--255.

\bibitem[{Galam et~al.(1998)Galam, Chopard, Masselot, and Droz}]{BC-galam:98}
Galam, S., Chopard, B., Masselot, A., and Droz, M. (1998).
\newblock Competing species dynamics: Qualitative advantage versus geography.
\newblock \emph{Eur. Phys. J. B}, 4, 529--531.

\bibitem[{Kanawati(2014)}]{kanawati2014}
Kanawati, R. (2014).
\newblock Seed-centric approaches for community detection in complex networks.
\newblock In \emph{International Conference on Social Computing and Social
  Media}, 197--208. Springer.

\bibitem[{Lin(1974)}]{Lin74}
Lin, C.T. (1974).
\newblock Structural controllability.
\newblock \emph{Automatic Control, IEEE Transactions on}, 19(3), 201--208.

\bibitem[{Liu et~al.(2011)Liu, Slotine, and Barabasi}]{Liu11}
Liu, Y.Y., Slotine, J.J., and Barabasi, A.L. (2011).
\newblock Controllability of complex networks.
\newblock \emph{Nature}, 473, 167.

\bibitem[{Liu et~al.(2013)Liu, Slotine, and Barab{\'a}si}]{Liu13}
Liu, Y.Y., Slotine, J.J., and Barab{\'a}si, A.L. (2013).
\newblock Observability of complex systems.
\newblock \emph{Proceedings of the National Academy of Sciences}, 110(7),
  2460--2465.

\bibitem[{Newman(2006)}]{newman2006}
Newman, M.E. (2006).
\newblock Modularity and community structure in networks.
\newblock \emph{Proceedings of the national academy of sciences}, 103(23),
  8577--8582.

\bibitem[{Ocampo-Mart{\'\i}nez et~al.(2011)Ocampo-Mart{\'\i}nez, Bovo, and
  Puig}]{ocampo2011}
Ocampo-Mart{\'\i}nez, C., Bovo, S., and Puig, V. (2011).
\newblock Partitioning approach oriented to the decentralised predictive
  control of large-scale systems.
\newblock \emph{Journal of Process Control}, 21(5), 775--786.

\bibitem[{Papadopoulos et~al.(2012)Papadopoulos, Kompatsiaris, Vakali, and
  Spyridonos}]{papadopoulos2012}
Papadopoulos, S., Kompatsiaris, Y., Vakali, A., and Spyridonos, P. (2012).
\newblock Community detection in social media.
\newblock \emph{Data Mining and Knowledge Discovery}, 24(3), 515--554.

\bibitem[{Schreiber(2000)}]{schreiber2000}
Schreiber, T. (2000).
\newblock Measuring information transfer.
\newblock \emph{Physical review letters}, 85(2), 461.

\bibitem[{Siljak(2011)}]{Siljak11}
Siljak, D.D. (2011).
\newblock \emph{Decentralized control of complex systems}.
\newblock Courier Corporation.

\bibitem[{Toupance(2019)}]{Toupance19}
Toupance, P-A \and L~Lef{\`e}vre, a.B.C. (2019).
\newblock Controllability of the voter model: an information theoretic
  approach.
\newblock \emph{ArXiv}.

\end{thebibliography}

\end{document}